\documentclass[twocolumn,secnumarabic,amssymb,nobibnotes,aps,prb,superscriptaddress]{revtex4-2}

\usepackage{graphicx}
\usepackage{hyperref}
\usepackage{ulem}
\hypersetup{
    colorlinks=true,
    citecolor=red,
    linkcolor=red,
    urlcolor=blue,
}
\usepackage{bm}
\usepackage{longtable}
\usepackage{amsmath,amssymb}
\usepackage{color}
\usepackage{empheq}
\usepackage{multirow}
\usepackage{float}
\usepackage[T1]{fontenc}
\newcommand{\figref}[1]{Fig.~\ref{#1}}
\newcommand{\tabref}[1]{Table~\ref{#1}}
\newcommand{\ave}[1]{\bigl\langle #1 \bigr\rangle}

\begin{document}

\title{Interband pairing in two-band superconductors with spin-orbit and Zeeman couplings}

\author{Shohei O. Shingu}
 \affiliation{Sapporo Shinyo High School, Hokkaido 005-0005, Japan}
\author{Jun Goryo}
 \affiliation{Department of Mathematics and Physics, Hirosaki University, Aomori 036-8561, Japan}

\date{\today}

% abstract
\begin{abstract}

Interband pairing in multiband superconductors is often neglected
because of its higher energetic cost compared with intraband pairing.
We show that, in multiband systems,
a Zeeman magnetic field can stabilize interband pairing
through the near degeneracy of spin-split branches from different bands,
even within a minimal on-site attractive interaction.
Using hexagonal tight-binding models with locally broken inversion symmetry,
we find a Zeeman-driven transition between
a conventional intraband $s$-wave state
and an interband-dominated superconducting Mixing state.
The resulting quasiparticle spectrum is intrinsically gapless,
leading to anomalous thermodynamic behavior,
including a $T$-linear specific heat at low temperatures,
reflecting a finite zero-energy density of states.

\end{abstract}

\maketitle

%========================= Introduction part =========================
%{\it \textcolor{blue}{Introduction.}}---%
\section{Introduction}

Multiband superconductors have attracted significant attention over the past decades.
Since the discovery of MgB$_2$ \cite{mgb2}, the importance of multiband and multiorbital
effects in superconductivity has been widely recognized.
From a theoretical viewpoint, the extension of the BCS theory to multiband systems was
originally explored by Suhl, Matthias, and Walker \cite{smw}.
In most theoretical studies, Cooper pairing within the same band
(intraband pairing) is assumed to dominate, 
while pairing between different bands (interband pairing) is often neglected
due to its higher energetic cost.
Nevertheless, interband pairing has been shown to play an essential role
in a variety of unconventional superconducting phenomena,
including Josephson tunneling, anomalous Hall responses,
gapless superconducting states,
and the stabilization of time-reversal-symmetry-breaking phases
\cite{joseph,halleff,gapless,ginz}.
Moreover, interband pairing can acquire nontrivial symmetry and topological character,
leading to nodal excitations or unconventional boundary modes
in multiband systems \cite{exotic,toporo}.

More generally, it has been shown that
interband pairing correlations emerge
in multiband superconductors
when band hybridization and inequivalent intra-band pairing amplitudes are present,
even when the pairing interaction itself is purely local
\cite{PhysRevB.88.104514}.
An external magnetic field can further enhance interband pairing
by bringing Zeeman-split branches of different bands closer in energy
\cite{highmag}.
These studies indicate that interband pairing is not an exotic possibility,
but a natural consequence of multiband electronic structure.

Despite these theoretical developments, explicit microscopic realizations
in which interband pairing becomes the dominant superconducting instability
remain relatively unexplored,
particularly in two-dimensional (2D) systems with locally broken inversion symmetry
and in relation to experimentally accessible thermodynamic signatures.
In this work, we address this issue by studying a minimal tight-binding model
on a 2D honeycomb lattice
that incorporates spin-orbit coupling (SOC) and a Zeeman magnetic field.
The honeycomb lattice provides a minimal two-sublattice structure
that naturally realizes a two-band electronic system.
Moreover, in the presence of intrinsic SOC,
it preserves spin-$S_z$ conservation,
allowing us to disentangle the effects of Zeeman splitting
and spin-orbit coupling in a transparent manner.
While the honeycomb geometry offers a particularly transparent realization,
the mechanism discussed below relies only on generic multiband structure
and Zeeman-induced near degeneracy.

Throughout this work, we consider a regime
where orbital pair-breaking effects are neglected
and the Zeeman effect dominates. 
Such a Zeeman-dominated situation can be realized in an ultrathin film
under an out-of-plane magnetic field,
provided that the sample size is much smaller than
the Pearl length $\lambda_{\mathrm{eff}} = \lambda_L^2/d$.
Here, $\lambda_L$ is the bulk London penetration depth
and $d$ is the film thickness.
Within this framework, we demonstrate that even a simple on-site attractive interaction
of electron--phonon origin can stabilize a nontrivial superconducting state
through the interplay of multiband structure, SOC, and Zeeman splitting.

Our main results are summarized as follows.
(i) The Zeeman field drives a transition between two distinct superconducting phases:
a conventional intraband $s$-wave state and a Mixing state in which
intraband and interband pairing components coexist.
(ii) SOC originating from local inversion-symmetry breaking
induces a characteristic mixing between intra- and interband pairing channels
in the Mixing state.
(iii) In the Mixing phase, the Bogoliubov quasiparticle spectrum becomes gapless
due to the combined effect of Zeeman splitting and the branch structure
of the quasiparticle bands.
As a consequence, the system exhibits anomalous thermodynamic behavior,
most notably a $T$-linear specific heat at low temperatures,
reflecting a finite residual density of states (DOS).

These features provide experimentally accessible fingerprints
of interband pairing in monolayer honeycomb superconductors,
clearly distinguishing the Mixing state
from a conventional fully gapped $s$-wave superconductor.

As an alternative platform, we also discuss
a three-dimensional (3D) non-symmorphic lattice model.
Since orbital magnetic effects are not included in our formulation,
the 3D case should be interpreted not as an electronic
superconductor in a bulk magnetic field,
but rather as a charge-neutral fermionic superfluid
in an artificial lattice, such as those realizable in ultracold atomic systems.
In such synthetic platforms, effective Zeeman fields can be implemented
without accompanying orbital coupling,
providing a clean setting to illustrate the generality
of the Zeeman-driven stabilization mechanism.

The remainder of this paper is organized as follows.
In Sec.~II, we introduce the model Hamiltonian.
Section~III is devoted to a detailed analysis of the superconducting mixing state, and 
the associated thermodynamic anomaly is discussed in Sec.~IV.
The summary and outlook are presented in Sec.~V.

% ========================= Model and Hamiltonian =========================
%{\it \textcolor{blue}{Model and Hamiltonian.}}---%

\section{Model}

\subsection{2D honeycomb model}

\begin{figure}%[H]
\includegraphics[width=\linewidth]{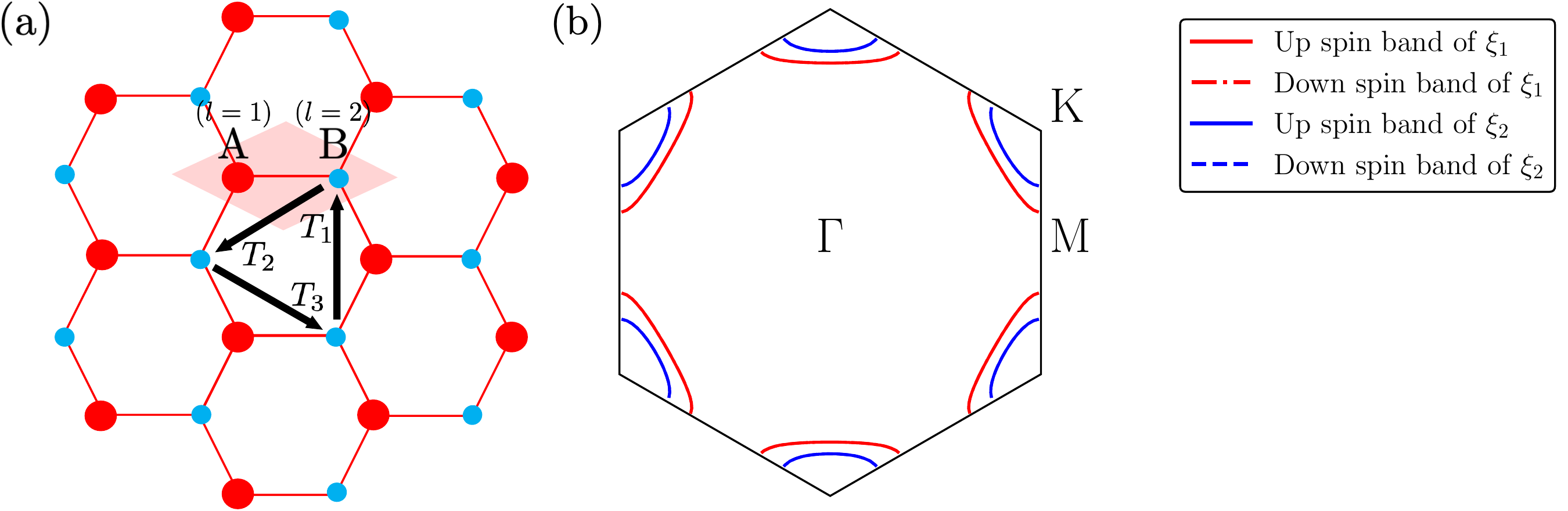}
\caption{\label{monomodel} Figure showing (a) the honeycomb lattice and (b) the example of Fermi surfaces without the SOC and the external magnetic field. }
%(a) Hexagonal monolayer model similar to graphene ( point group $D_{6 h}$ or $6/mmm$ ). (b) The example of Fermi surfaces without the SOC and the external magnetic field.}
\end{figure}

We first introduce a minimal 2D tight-binding model defined on a honeycomb lattice (see Fig.~\ref{monomodel}), which serves as the primary platform of this study.
The unit cell consists of two sublattices, labeled A and B, and the lattice lacks local inversion symmetry, 
giving rise to spin-orbit coupling (SOC).
Such a setting naturally describes monolayer superconductors with multiple electronic degrees of freedom and locally broken inversion symmetry.

The normal-state Hamiltonian is written as
\begin{align}
\label{hamham}
H_{0} &= \sum_{\bm{k},\sigma} 
\left(
\begin{array}{cc}
c^{\dag}_{1 \bm{k} \sigma} & c^{\dag}_{2 \bm{k} \sigma}
\end{array}
\right)
\hat{h}_{0\bm k \sigma}
\left(
\begin{array}{c}
c_{1 \bm{k} \sigma} \\
c_{2 \bm{k} \sigma}
\end{array}
\right),
\\
\hat{h}_{0\bm k \sigma}&=
\left(
\begin{array}{cc}
\epsilon_{\bm{k}} + ( \hat{\sigma}_{z} )_{\sigma \sigma} \, \alpha \lambda_{\bm{k}} & \epsilon_{c\bm{k}} \\
\epsilon^{*}_{c\bm{k}} & \epsilon_{\bm{k}} - ( \hat{\sigma}_{z} )_{\sigma \sigma} \, \alpha \lambda_{\bm{k}}
\end{array}
\right), 
\end{align}
where $c_{l\bm{k}\sigma}$ ($c^{\dag}_{l\bm{k}\sigma}$) annihilates (creates) an electron with momentum $\bm{k}$ and spin $\sigma$ on sublattice $l=1,2$.
The kinetic energy is given by
$\epsilon_{\bm{k}} = - t \sum^{3}_{i=1} \cos(\bm{k} \cdot \bm{T}_{i}) - \mu$,
and the SOC form factor is
$\lambda_{\bm{k}} = \sum_{i = 1}^{3} \sin(\bm{k} \cdot \bm{T}_{i})$,
with in-plane nearest-neighbor bond vectors
$\bm{T}_{1} = (0,a,0)$,
$\bm{T}_{2} = (-\sqrt{3}a/2,-a/2,0)$,
and
$\bm{T}_{3} = (\sqrt{3}a/2,-a/2,0)$.
The inter-sublattice hybridization
$\epsilon_{c\bm{k}} = -t_c(1+e^{i \bm k \cdot \bm T_2}+e^{-i \bm k \cdot \bm T_3})=|\epsilon_{c\bm{k}}| e^{i\theta_{\bm{k}}}$
satisfies $\theta_{\bm{k}} = - \theta_{-\bm{k}}$ due to
$\epsilon_{c\bm{k}} = \epsilon^{*}_{c,-\bm{k}}$.
Here $\hat{\sigma}_{i}$ denotes the Pauli matrices in spin space.
In the absence of $\epsilon_{\bm{k}}$, the Hamiltonian $\hat{h}_{0\bm k \sigma}$
coincides with the Kane--Mele model~\cite{PhysRevLett.95.146802} for a honeycomb lattice
with intrinsic spin-orbit coupling.

An external magnetic field is introduced through the Zeeman (Pauli) coupling,
\begin{align}
H_{\rm mag}
= h \sum_{l,\bm{k},\sigma}
( \hat{\sigma}_{z} )_{\sigma \sigma}
c^{\dag}_{l \bm{k} \sigma}
c_{l \bm{k} \sigma}.
\end{align}
Throughout this work, we neglect orbital coupling to the vector potential and focus exclusively on the Zeeman effect.
This approximation is well justified in ultrathin 2D systems, as mentioned. 

The Hamiltonian $H_{0}+H_{\rm mag}$ is diagonalized in the band basis via the unitary transformation
$c_{l \bm{k} \sigma} = \sum_{b} ( \hat{U}_{\bm{k}\sigma} )_{lb} \, a_{b \bm{k} \sigma}$.
The resulting normal-state energy spectrum is
$\xi_{b\bm{k}} + ( \hat{\sigma}_{z} )_{\sigma \sigma} h$,
where
$\xi_{b\bm{k}} = \epsilon_{\bm{k}} + (-1)^{b-1} C_{\bm{k}}$
and
$C_{\bm{k}} = \sqrt{\alpha^{2} \lambda^{2}_{\bm{k}} + |\epsilon_{c\bm{k}}|^{2}}$.

%%%%%%%%%%%%%%%%%%
%%%%%%%%%%%%%%%%%%

\subsection{Extension to a 3D non-symmorphic model}

\begin{figure}%[H]
 \includegraphics[width=\linewidth]{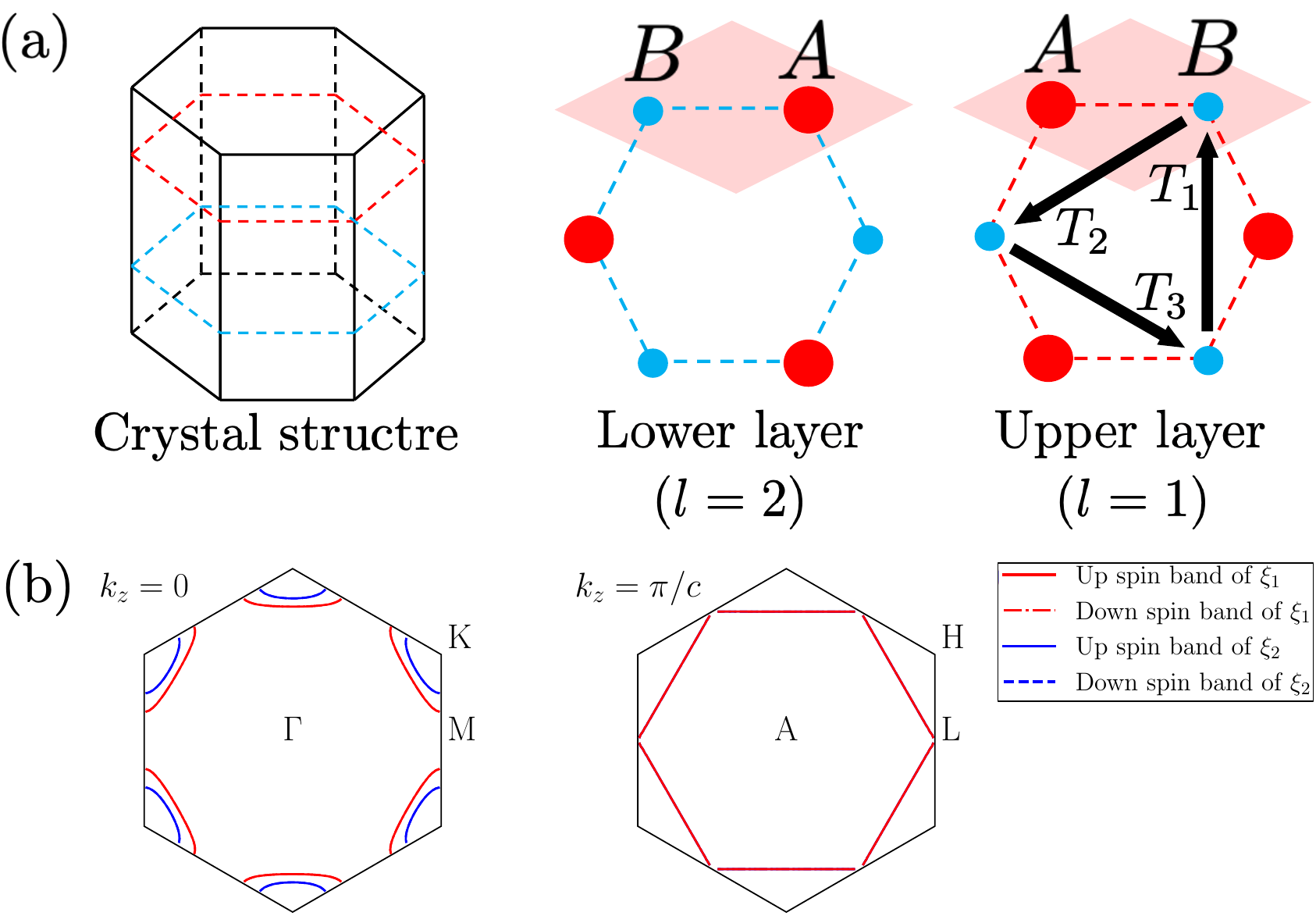}
 \caption{\label{3dmodelfer} Figure showing lattice and Fermi surfaces. (a) Non-symmorphic two-layer hexagonal model similar to SrPtAs ( space group $D^{4}_{6 h}$ or ${\it P} 6_{3} / mmc$ ). (b) The example of Fermi surfaces without the SOC and the external magnetic field. There is a band degeneracy at $k_{z} = \pi / c$.}
\end{figure}

As a minimal extension of the 2D model introduced above,
we also consider a 3D non-symmorphic two-layer hexagonal model
similar to SrPtAs \cite{srptas}, illustrated in Fig.~\ref{3dmodelfer}.
The unit cell contains two hexagonal layers rotated by $\pi/3$ relative to each other, such that the crystal as a whole preserves inversion symmetry, while each layer individually lacks an inversion center.
This structure gives rise to layer-dependent SOC with opposite signs.
In the 3D model, an electron hops on the $A$ sublattice cites, and $l=1,2$ denotes the layer index. Also,  
$\epsilon_{\bm{k}} = - t \sum^{3}_{i=1} \cos(\bm{k} \cdot \bm{T}_{i}) - t_{c2} \cos k_z c - \mu$ and 
$\epsilon_{c\bm{k}} =- t_c(1+e^{i \bm k \cdot \bm T_2}+e^{-i \bm k \cdot \bm T_3}) \cos{k_z c/2}$. 

In the absence of SOC and an external magnetic field,
the non-symmorphic symmetry enforces band degeneracy
at the Brillouin-zone boundary ($k_{z}=\pi/c$),
which provides a natural condition for interband pairing
with zero center-of-mass momentum.
When SOC and Zeeman splitting are introduced,
this degeneracy is generally lifted,
leading to an energy mismatch between the bands.
However, as discussed in Ref.~\cite{highmag},
a Zeeman field can, under appropriate conditions,
induce near degeneracy between spin-split branches
from different bands,
thereby enhancing interband pairing.

In the limit $t_{c2}\to 0$ and $k_{z}\to 0$, the 3D model smoothly reduces to the 2D honeycomb model introduced above.
While the 3D case is less directly applicable to electronic superconductors under magnetic fields, it can be interpreted as a minimal description of a charge-neutral fermionic superfluid in an optical lattice subject to an effective Zeeman field.
This extension highlights the generality of the Zeeman-dominated mechanism discussed in this work.

%%%%%%%%%%%%%%%%%%
%%%%%%%%%%%%%%%%%%

% ========================= Superconducting state =========================
%\textcolor{blue}{Superconducting state:\enspace}%
%%%%%%%%%
%%%%%%%%%
%%%%%%%%%
%%%%%%%%%
\section{Superconducting state}

We now investigate the superconducting instability of the models 
introduced in Sec.~II.
As a minimal interaction that captures the essential pairing channels in multiband systems,
we consider an on-site interaction in the 2D model (at each layer's A site in the 3D model). 
Although the interaction is local in real space, its projection onto the band basis
naturally generates multiple pairing channels due to the multiband structure and SOC.
The Hamiltonian for the pairing interaction $H_{int}$ is as follows:
\begin{align}
\label{hint}
H_{int} & = - \frac{g}{2 N^{d}} \sum_{l,\bm{k},\bm{k}',\sigma} 
c^{\dag}_{l \bm{k} \sigma} c^{\dag}_{l -\bm{k} -\sigma} c_{l - \bm{k}' -\sigma} c_{l \bm{k}' \sigma} %\\
%& \Bigl[ c^{\dag}_{l \bm{k} \uparrow} c^{\dag}_{l -\bm{k} \downarrow} c_{l - \bm{k}' \downarrow} c_{l \bm{k}' \uparrow} + c^{\dag}_{l \bm{k} \downarrow} c^{\dag}_{l -\bm{k} \uparrow} c_{l - \bm{k}' \uparrow} c_{l \bm{k}' \downarrow} \Bigr] \notag
\end{align}
where $N^{d}$ represents the total number of unit cells, and $d = 2 (3)$ in the hexagonal monolayer model 
(non-symmorphic two-layer hexagonal model).
The symbol $-\sigma$ corresponds to anti-spin of $\sigma$. For example, if $\sigma = \uparrow$, then $-\sigma = \downarrow$.
By employing the unitary transformation on $H_{int}$ and applying the mean-field approximation 
in the band space, we obtain the superconducting state Hamiltonian:

\begin{align}
\label{surperham}
H & = \sum_{\bm{k},\sigma} \Psi^{\sigma \, \dag}_{\bm{k}} \, \overline{H}^{\sigma}_{\bm{k}} \, \Psi^{\sigma}_{\bm{k}} + \frac{1}{2} \sum_{\bm{k},b,\sigma} \Bigl\{ \xi_{b \bm{k}} + ( \hat{\sigma}_{z} )_{\sigma \sigma} h \Bigr\} + N^{d} W
\end{align}
\begin{align}
\overline{H}^{\sigma}_{\bm{k}} & = 
\left(
\begin{array}{cc}
\hat{\xi}_{\bm{k} \sigma}                      & \hat{\Delta}^{\sigma \, -\sigma}_{\bm{k}} \\
\hat{\Delta}^{\sigma \, -\sigma \, \dag}_{\bm{k}} & - \hat{\xi}_{\bm{k} -\sigma}
\end{array}
\right)
\end{align}
\begin{align}
\hat{\xi}_{\bm{k} \sigma} & =
\left(
\begin{array}{cc}
\xi_{1 \bm{k}} + ( \hat{\sigma}_{z} )_{\sigma \sigma} h &                \\
                                                        & \xi_{2 \bm{k}} + ( \hat{\sigma}_{z} )_{\sigma \sigma} h
\end{array}
\right)
\end{align}
\begin{align}
\label{surperorder}
& \hat{\Delta}^{\sigma \, -\sigma}_{\bm{k}} = 
\left(
\begin{array}{cc}
\Delta^{\sigma \, -\sigma}_{11}(\bm{k}) & \Delta^{\sigma \, -\sigma}_{12}(\bm{k}) \\
\Delta^{\sigma \, -\sigma}_{21}(\bm{k}) & \Delta^{\sigma \, -\sigma}_{22}(\bm{k})
\end{array}
\right)
\end{align}
where $\Psi^{\sigma}_{\bm{k}} = ( a_{1 \bm{k} \sigma}, a_{2 \bm{k} \sigma}, a_{1 -\bm{k} -\sigma}^{\dag}, a_{2 -\bm{k} -\sigma}^{\dag} )^{T} / \sqrt{2}$.
The superconducting order parameters $\Delta^{\sigma \, -\sigma}_{11}(\bm{k})$ and $\Delta^{\sigma \, -\sigma}_{22}(\bm{k})$ represent the intraband pairing, while $\Delta^{\sigma \, -\sigma}_{12}(\bm{k})$ and $\Delta^{\sigma \, -\sigma}_{21}(\bm{k})$ represent {\it the interband pairing}.
%These states are characterized by $\Delta_{0}$ and $\Delta_{1}$.

At the mean-field level, the superconducting order parameter in the band space
can be decomposed into two symmetry-distinct components.
The first, $\Delta_0$, corresponds to a conventional intraband spin-singlet pairing channel,
while the second, $\Delta_1$, represents a mixing channel that intrinsically involves interband pairing.
As shown below, these two components decouple at the linearized level,
which allows the superconducting instability to be classified into two independent types.
$\Delta_0$ and $\Delta_1$ are defined as: 
\begin{eqnarray}
\Delta_{0}
 &=& - \frac{g}{2 N^{d}} \sum_{\bm{k},m} \ave{a_{m -\bm{k} \downarrow} \ a_{m \bm{k} \uparrow}},  
 \\
\Delta_{1}
&=& - \frac{g}{2 N^{d}} \sum_{\bm{k},m,m'} 
 \frac{\ave{a_{m -\bm{k} \downarrow} \ a_{m' \bm{k} \uparrow}} }{C_{\bm{k}}}
 \\
&& \times  \Bigl\{ \text{Re} \, \epsilon_{c \bm{k}} \, \hat{\tau}_{1} - \text{Im} \, \epsilon_{c \bm{k}} \, \hat{\tau}_{2} + \alpha \lambda_{\bm{k}} \hat{\tau}_{3} \Bigr\}_{m m'} \notag,  
\end{eqnarray}
%\vspace{-0.6cm} %
where $\hat{\tau}_{i}$ is the Pauli matrix with band indices $m$ and $m'$. 
These order parameters correspond to distinct pairing channels allowed by 
the on-site interaction in the band representation.
Using $\Delta_{0}$ and $\Delta_{1}$, 
\begin{align}
\Delta^{\sigma \, -\sigma}_{m m}(\bm{k}) & = i ( \hat{\sigma}_{y} )_{\sigma \, -\sigma} ( \hat{\tau}_{0} )_{m m} \, \Delta_{0} + \frac{( \hat{\sigma}_{x} )_{\sigma -\sigma} \, ( \hat{\tau}_{3} )_{m m} \, \alpha \lambda_{\bm{k}}}{C_{\bm{k}}} \Delta_{1} \\
\Delta^{\sigma \, -\sigma}_{m m'}(\bm{k}) & = i ( \hat{\sigma}_{y} )_{\sigma \, -\sigma} \, \frac{ \Bigl\{ \text{Re} \, \epsilon_{c \bm{k}} \, \hat{\tau}_{1} - \text{Im} \, \epsilon_{c \bm{k}} \, \hat{\tau}_{2} \Bigr\}_{m m'} }{C_{\bm{k}}} \Delta_{1}
\end{align}

%\begin{align}
%\Delta^{\sigma \, -\sigma}_{11}(\bm{k}) & = i ( \hat{\sigma}_{y} )_{\sigma \, -\sigma} \, \Delta_{0} + \frac{( \hat{\sigma}_{x} )_{\sigma -\sigma} \, \alpha \lambda_{\bm{k}}}{C_{\bm{k}}} \Delta_{1}
%\end{align}
%\begin{align}
%\Delta^{\sigma \, -\sigma}_{12}(\bm{k}) & = i ( \hat{\sigma}_{y} )_{\sigma \, -\sigma} \, \frac{\epsilon_{c\bm{k}}}{C_{\bm{k}}} \Delta_{1}
%\end{align}
%\begin{align}
%\Delta^{\sigma \, -\sigma}_{21}(\bm{k}) & = i ( \hat{\sigma}_{y} )_{\sigma \, -\sigma} \, \frac{\epsilon_{c\bm{k}}^{*}}{C_{\bm{k}}} \Delta_{1}
%\end{align}
%\begin{align}
%\Delta^{\sigma \, -\sigma}_{22}(\bm{k}) & = i ( \hat{\sigma}_{y} )_{\sigma \, -\sigma} \, \Delta_{0} - \frac{( \hat{\sigma}_{x} )_{\sigma -\sigma} \, \alpha \lambda_{\bm{k}}}{C_{\bm{k}}} \Delta_{1}
%\end{align}
\begin{align}
W & = \frac{2 ( |\Delta_{0}|^{2} + |\Delta_{1}|^{2} )}{g}
\end{align}
and the eigenvalues of $\overline{H}^{\sigma}_{\bm{k}}$ are 
$\pm E_{m \bm{k}} + (\hat{\sigma}_{z})_{\sigma \sigma}h$, where
\begin{align}
E_{m \bm{k}} 
  & = \sqrt{A + (-1)^{m-1} 2 \sqrt{B}} 
  \label{QP_spectrum}\\
A & = |\Delta_{0}|^{2} + |\Delta_{1}|^{2} + \epsilon_{\bm{k}}^{2} + C^{2}_{\bm{k}} 
\nonumber\\
B & = |\Delta_{0}|^{2} |\Delta_{1}|^{2} \cos^{2}{\theta_{1}} + 2 \alpha \lambda_{\bm{k}} \epsilon_{\bm{k}} |\Delta_{0}| |\Delta_{1}| \cos{\theta_{1}} 
\nonumber\\
  & + |\epsilon_{c \bm{k}}|^{2} |\Delta_{1}|^{2} + \epsilon_{\bm{k}}^{2} C_{\bm{k}}^{2},  
\nonumber
\end{align}
and $\theta_1 = \arg(\Delta_0) - \arg(\Delta_1)$ 
denotes the relative phase between the two order parameters. 

To obtain the transition temperature $T_c$, we examine the linearized gap equation: 
\begin{align}
\label{del0tc}
\Delta_{0} & = Sum(1) \Delta_{0} + Sum(3) \Delta_{1} \\
\label{del1tc}
\Delta_{1} & = Sum(3) \Delta_{0} + Sum(2) \Delta_{1}
\end{align}
where
\begin{align}
& Sum(1) = \frac{g}{4 N^{d}} \sum_{\bm{k},m} \frac{1}{\xi_{m \bm{k}}} \tanh^{T_{c}}_{m} \\
& Sum(2) = \frac{g}{4 N^{d}} \sum_{\bm{k},m} \frac{1}{C_{\bm{k}}^{2}} \Biggl\{ \frac{|\epsilon_{c \bm{k}}|^{2}}{\epsilon_{\bm{k}}} + \frac{\alpha^{2} \lambda_{\bm{k}}^{2}}{\xi_{m \bm{k}}} \Biggr\} \tanh^{T_{c}}_{m} \\
& Sum(3) = \frac{g}{4 N^{d}} \sum_{\bm{k},m} 
 \frac{\alpha \lambda_{\bm{k}}}{C_{\bm{k}}} \frac{(-1)^{m-1}}{\xi_{m \bm{k}}} \tanh^{T_{c}}_{m} \\
& \tanh^{T_{c}}_{m} = \tanh{ \frac{ \xi_{m \bm{k}} + h }{2 k_{B} T_{c}} } + \tanh{ \frac{ \xi_{m \bm{k}} - h }{2 k_{B} T_{c}} }
\end{align}
However, $Sum(3)$ would vanish, since its integrand is odd with respect to $\bm{k}$.
As a consequence, $\Delta_{0}$ and $\Delta_{1}$ become completely independent
at the linearized level.
This justifies the introduction of $\Delta_{0}$ and $\Delta_{1}$ as separate
order parameters and shows that the superconducting instability splits into
two distinct types.
The $4 \times 4$ gap matrix (\ref{surperorder}) in each case is
\begin{eqnarray}
\hat{\Delta}_{s \bm{k}}&=&\Delta_0 \psi^{0}_{\bm{k}} \ i \hat{\sigma}_{y} \otimes \hat{\tau}_{0}
\label{intrapair}
\\
\hat{\Delta}_{M \bm{k}}&=&\Delta_1(\psi^{1}_{\bm{k}}\ i \hat{\sigma}_{y} \otimes \hat{\tau}_{1} + \psi^{2}_{\bm{k}}\ i \hat{\sigma}_{y} \otimes \hat{\tau}_{2} 
\label{mixingpair}\\
&&+ d^{3}_{z \bm{k}}\ i \hat{\sigma}_{z}\hat{\sigma}_{y} \otimes \hat{\tau}_{3}) 
\nonumber
\end{eqnarray}
with basis functions listed in \tabref{tabtau} and \tabref{sympari}. 
These states satisfy the Pauli principle $\hat{\Delta}(\bm{k}) = - \hat{\Delta}^{T}(-\bm{k})$.
The on-site pairing interaction in Eq.\eqref{hint} creates pairing with $S_{z} =0$ ($S_{z}; z$ component of spin).
%Therefore, the spin parts are represented by the spin-singlet part $i \hat{\sigma}_{y}$ 
%or a linear combination of the singlet- and spin-triplet part 
%$i (\hat{\bm{\sigma}})_{z} \hat{\sigma}_{y} = \hat{\sigma}_{x}$.
%In fact, Eqs.\eqref{intrapair} and \eqref{mixingpair} can be expressed in this manner.
Eq.\eqref{intrapair} represents a conventional intraband $s$-wave pairing.
On the other hand, Eq.\eqref{mixingpair} represents a mixing of the intraband spin-triplet $d^{3}_{z \bm{k}}$ and the interband spin-singlet pairings $\psi^{1}_{\bm{k}}$ and $\psi^{2}_{\bm{k}}$ with even and odd parity.
The intraband spin-triplet pairing $d^{3}_{z \bm{k}}$ is proportional to $\alpha$ and caused by the SOC with local inversion symmetry breaking.

\begin{figure*}%[H]
 \includegraphics[width=\linewidth]{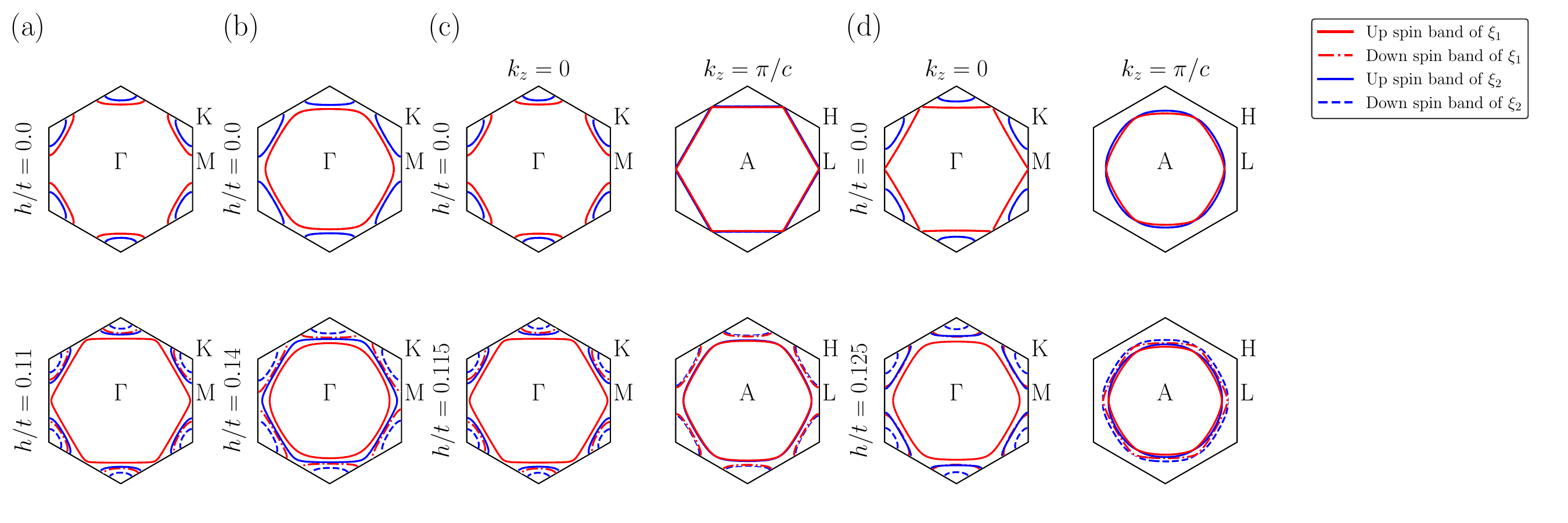}
 \caption{\label{fermis} The Fermi surfaces. 
(a) 2DFS1 with $h / t = 0.0$, $h / t = 0.11$. (b) 2DFS2 with $h / t = 0.0$, $h / t = 0.14$. 
 (c) 3DFS1 with $h / t = 0.0$, $h / t = 0.115$. (d) 3DFS2 with $h / t = 0.0$, $h / t = 0.125$. 
 The external magnetic field $h$ reduces the energy mismatch 
 at the Fermi level between the down-spin branch of band $\xi_{\bm k 1}$
 and the up-spin branch of band $\xi_{\bm k2}$.}
\end{figure*}

To investigate the stability of Mixing state with interband pairing, we introduce four types of Fermi surfaces refferd to as 
2DFS1 and 2DFS2 for the honeycomb monolayer model, and 3DFS1 and 3DFS2 for the non-symmorphic two-layer model
(see \tabref{ferpara} and \figref{fermis}).
We compare the transition temperatures of the pairing states by numerically 
solving Eqs.~\eqref{del0tc} and \eqref{del1tc} using these Fermi surfaces.
In the following, the value of $g / t$ is set to $0.5$.

\begin{figure}%[H]
 \includegraphics[width=\linewidth]{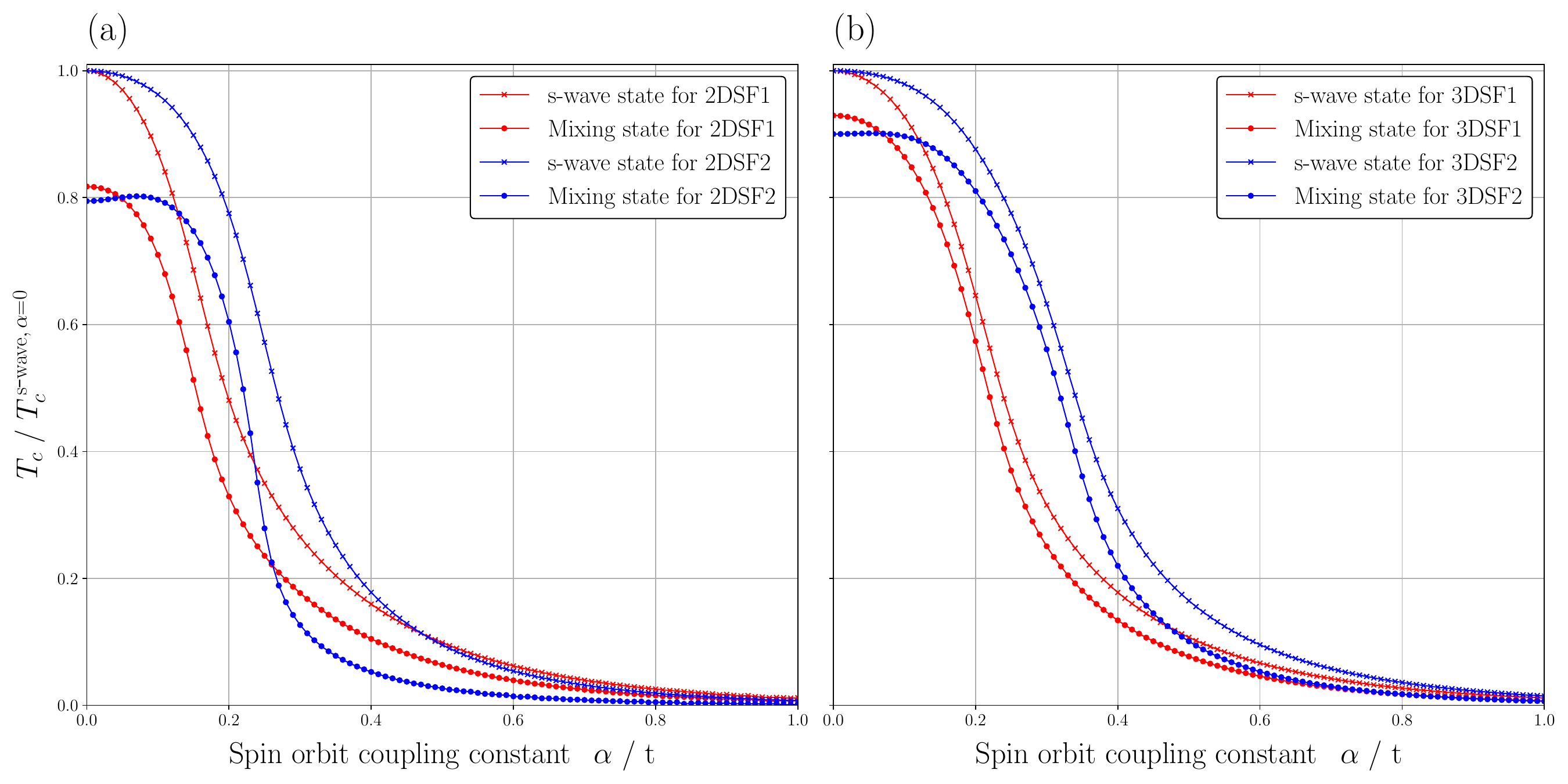}
 \caption{
 \label{kbtcsoc} 
 (a) Transition temperature $T_c$ as a function of the SOC strength $\alpha$ 
for the 2D model at $h=0$.
The red (blue) curves correspond to the tight-binding parameters 
of 2DFS1 (2DFS2) listed in Table~\ref{ferpara},
with all parameters fixed except for $\alpha$.
(b) Corresponding results for the 3D models.
The red (blue) curves represent the parameter sets 
3DFS1 (3DFS2) in Table~\ref{ferpara},
again varying only $\alpha$.
In both dimensions, the transition temperature of the $s$-wave state 
is higher than that of the Mixing state.
}
 %\end{minipage}
 \end{figure}

\begin{figure}%[H]
  \includegraphics[width=\linewidth]{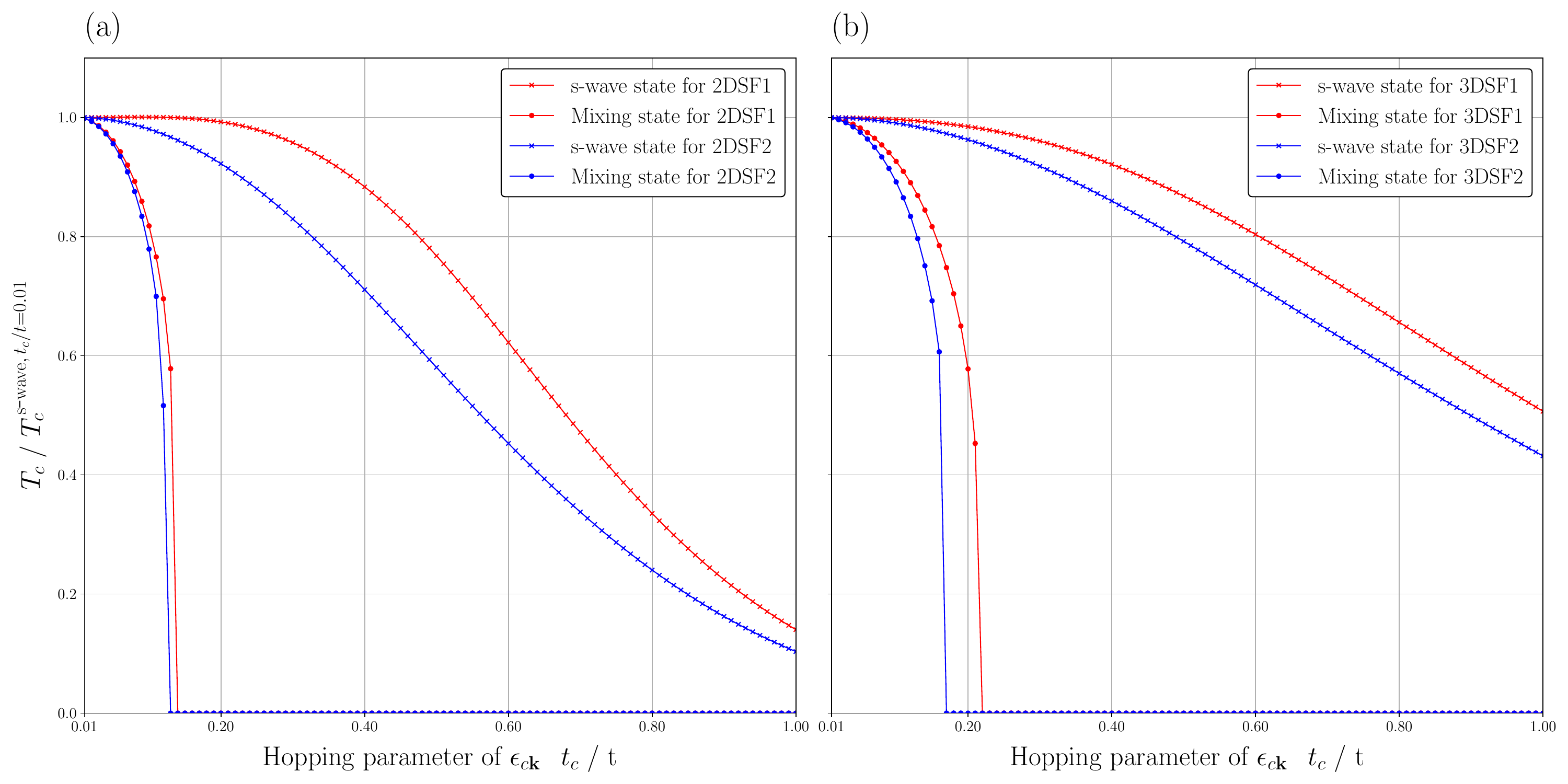}
 \caption{\label{kbtctc} 
(a) Transition temperature $T_c$ as a function of the hopping parameter $t_c$
for the 2D model at $h=\alpha=0$.
The red (blue) curves correspond to the parameter sets
2DFS1 (2DFS2) listed in Table~\ref{ferpara},
with all parameters fixed to those values except that
$\alpha$ is set to zero and $t_c$ is varied.
(b) Corresponding results for the 3D models.
The red (blue) curves represent the parameter sets
3DFS1 (3DFS2) in Table~\ref{ferpara},
again with $\alpha=0$ and varying only $t_c$.
For $\alpha=0$, the Mixing state reduces to purely interband pairing.
In both dimensions, the transition temperature of the $s$-wave state
is higher than that of the Mixing state.
}
%\end{minipage}
 \end{figure}

We find that the transition temperature of the conventional $s$-wave state
is consistently higher than that of the Mixing state,
both in the presence of SOC (Fig.~\ref{kbtcsoc})
and in the case where only the hybridization term $\epsilon_{c\bm{k}}$ is present
(Fig.~\ref{kbtctc}).
This can be attributed to the fact that increasing SOC or $\epsilon_{c \bm{k}}$ enlarges the band separation 2 $C_{\bm{k}}$ without spin splitting, the Mixing state is not favored. 
Although $\epsilon_{c\bm{k}}$ is required to generate interband pairing components
(since $\Delta_{12}(\bm{k})$ and $\Delta_{21}(\bm{k})$ are proportional to $\epsilon_{c\bm{k}}$),
our numerical results show that neither SOC nor $\epsilon_{c\bm{k}}$
significantly enhances the stability of the Mixing state.

\begin{figure}%[H]
 \includegraphics[width=\linewidth]{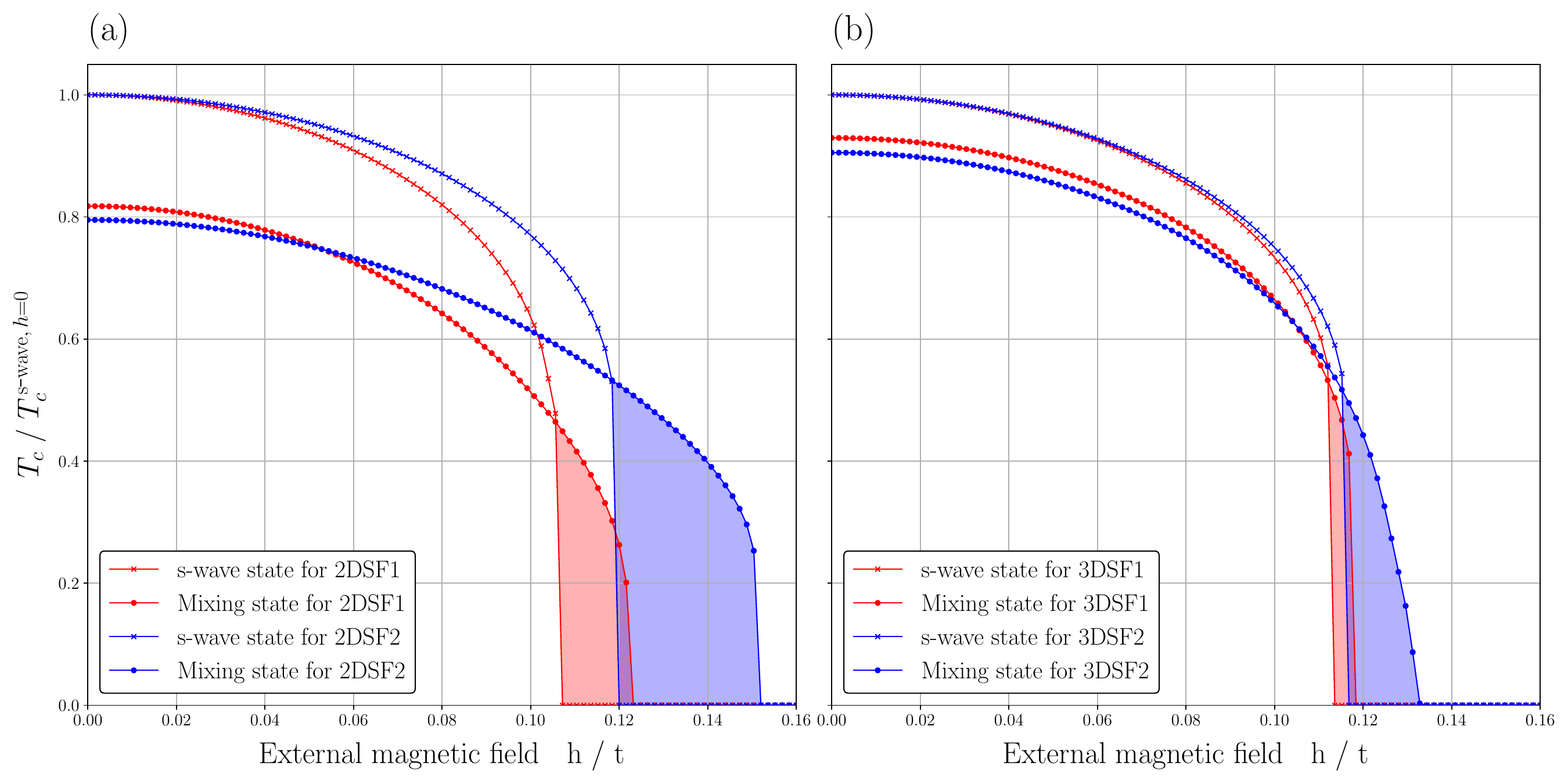}
 \caption{\label{kbtch} Figure showing results of $T_{c}$ vs $h$ 
 for (a) 2D, and (b) 3D models. 
 In both phase diagram, the color region shows the dominance of the Mixing state. The blue region is wider than the red region, indicating that 2DFS2 (3DFS2) is favorable for interband pairing in the 2D (3D) case.}
 \end{figure}

We next examine the role of an external magnetic field,
which provides the key mechanism for stabilizing interband pairing.
The results shown in~\figref{kbtch} indicate that, at low magnetic fields,
the conventional $s$-wave state is favored over the Mixing state
in both 2D and 3D models.
As the magnetic field approaches the Pauli-limiting field
associated with the conventional $s$-wave state,
its transition temperature is strongly suppressed.
Beyond this scale, the Mixing state can become energetically favorable.
In this sense, the boundary between the low- and high-field regimes
is naturally interpreted in terms of the Pauli limit of the intraband pairing.

However, exceeding the Pauli limit alone is not sufficient
to guarantee the emergence of the Mixing state.
The interband-dominated regime appears only when the Zeeman splitting
brings the spin-down branch originating from one band
into near degeneracy with the spin-up branch of the other band.
If this near-degeneracy condition is not realized,
the Mixing state does not develop even above the Pauli-limiting field.

Thus, the Zeeman field plays a dual role:
it suppresses the conventional intraband pairing,
and, when the band structure is favorable,
it promotes interband pairing by creating the required near degeneracy.
This mechanism ultimately determines the extent of the Mixing-state region
in the magnetic-field phase diagram.

In particular, for the 2DFS2 [\figref{fermis}(b)] and 3DFS2 [\figref{fermis}(d)] 
Fermi surfaces, these Zeeman-split branches become nearly degenerate at 
the Fermi level, while the Fermi level itself lies 
in proximity to the Van Hove singularity. As a consequence, 
the enhanced DOS near the Fermi level, together with the reduced energy 
mismatch between the interband pairing partners, leads to an increase in the transition 
temperature of the Mixing state in both models.

\begin{table}%[t]
 \caption{\label{tabtau} Classification of the Pauli matrices $\hat{\tau}_{a}$ in the band space. Parity arises from exchanging the band index.}
 \begin{ruledtabular}
 \begin{tabular}{ccccc}
 & Parity & Intraband                          & Interband        &  \\ \hline
 & Even   & $\hat{\tau}_{0}, \ \hat{\tau}_{3}$ & $\hat{\tau}_{1}$ &  \\ 
 & Odd    &                                    & $\hat{\tau}_{2}$ &  
 \end{tabular}
 \end{ruledtabular}
 \caption{\label{sympari} Classification of basis functions $\psi^{i}_{\bm{k}}$ and $d^{3}_{z \bm{k}}$. 
 Parity arises from exchanging $\bm{k}$ to $-\bm{k}$.}
 \begin{ruledtabular}
 \begin{tabular}{ccccc}
 & Parity                &      Spin-Singlet                                                                                  & Spin-Triplet  & \\ \hline
 & \multirow{2}{*}{Even} & $ \psi^{0}_{\bm{k}} = 1 $                                                                 &               & \\
 &                       & $ \psi^{1}_{\bm{k}} = \frac{|\epsilon_{c\bm{k}}| }{C_{\bm{k}}} \cos{\theta_{\bm{k}}} $  &               & \\ \hline
 & Odd                   & $ \psi^{2}_{\bm{k}} = - \frac{|\epsilon_{c\bm{k}}| }{C_{\bm{k}}} 
 \sin{\theta_{\bm{k}}}$ & $d^{3}_{z \bm{k}} = \frac{\alpha \lambda_{\bm{k}}}{C_{\bm{k}}} $ & 
 \end{tabular}
 \end{ruledtabular}
 \caption{\label{ferpara}  Tight-binding parameters. }
 \begin{ruledtabular}
 \begin{tabular}{cccccccc}
 & Fermi surface & $t_{c} / t$ & $t_{c2} / t$ & $\mu / t$ & $\alpha / t$ & $c / a$ & \\ \hline
 & 2DFS1         & 0.1         &              & 1.2       & 0.01         &         & \\ 
 & 2DFS2         & 0.1         &              & 1.0       & 0.01         &         & \\  \hline
 & 3DFS1         & 0.1         & 0.1          & 1.1       & 0.01         & 1.0     & \\
 & 3DFS2         & 0.1         & 0.2          & 0.9       & 0.05         & 1.0     & 
 \end{tabular}
 \end{ruledtabular}
\end{table}
%

%%%%%%%%%%%%%%%%%%%%%%%%%%%%%%
%%%%%%%%%%%%%%%%%%%%%%%%%%%%%%

%%%%%%%%%%%%%%%%%%%%%%%%%%%%%%
%\begin{figure*}%[H]
%\begin{minipage}[b]{0.49\linewidth}
 %\end{figure*}
 %%%%%%%%%%%%%%%%%%%%%%%%%%%%%%
 
 %%%%%%%%%%%%%%%%%%%%%%%%%%%%%%
%\begin{figure*}%[H]
%\begin{minipage}[b]{0.49\linewidth}
 %\end{figure*}
%%%%%%%%%%%%%%%%%%%%%%%%%%%%%%

%

%%%%%%%%%%%%%%%%%%%%%%%%%%%%%%
%\begin{figure*}%[H]
 %\end{figure*}
%%%%%%%%%%%%%%%%%%%%%%%%%%%%%%

%========================= Thermodynamic property part =========================
\section{Thermodynamic properties}

We now examine thermodynamic signatures of the superconducting states,
focusing on experimentally accessible quantities.
In particular, we show that the Mixing state exhibits anomalous low-temperature
behavior originating from low-energy quasiparticles,
in sharp contrast to a fully gapped conventional $s$-wave state.
To study the behavior of these states at finite temperatures, we solve the gap equation self-consistently: 
\begin{align}
\label{gapes}
\Delta_{\gamma} & = \frac{g}{4 N^{d}} \sum_{\bm{k},m} \frac{\Delta_{\gamma}}{E_{m \bm{k}}} D_{\gamma} \tanh^{T}_{m} \\
D_{\gamma} = & \left\{
\begin{array}{ll}
1 & \ (\gamma = 0:\text{ for $s$-wave state }) \\
1 + \frac{(-1)^{m-1} |\epsilon_{c \bm{k}}|^{2} }{\sqrt{B}} & \ (\gamma = 1:\text{ for Mixing state })
\end{array}
\right. \\
& \tanh^{T}_{m} = \tanh{\frac{E_{m \bm{k}} + h}{2 k_{B} T}} + \tanh{\frac{E_{m \bm{k}} - h}{2 k_{B} T}}
\end{align}
where $A$ and $B$ in Eq. (\ref{QP_spectrum})
%$E_{m \bm{k}}=\sqrt{A + (-1)^{m-1} \sqrt{B}} $ (see Eq. (\ref{QP_spectrum})) 
are 
\begin{align}
A & = \left\{
\begin{array}{ll}
\xi_{m \bm{k}}^{2} + \Delta_{0}^{2} & \text{ for $s$-wave state } \\
\epsilon_{\bm{k}}^{2} + C_{\bm{k}}^{2} + \Delta_{1}^{2} & \text{ for Mixing state }
\end{array}
\right.\\
B & = \left\{
\begin{array}{ll}
0 & \text{ for  $s$-wave state } \\
|\epsilon_{c \bm{k}}|^{2} \Delta_{1}^{2} + \epsilon_{\bm{k}}^{2} C_{\bm{k}}^{2} & \text{ for Mixing state }
\end{array}
\right.
\end{align}
and satisfies $A\geq 2\sqrt{B}$. 

\begin{figure*}
 \includegraphics[width=\linewidth]{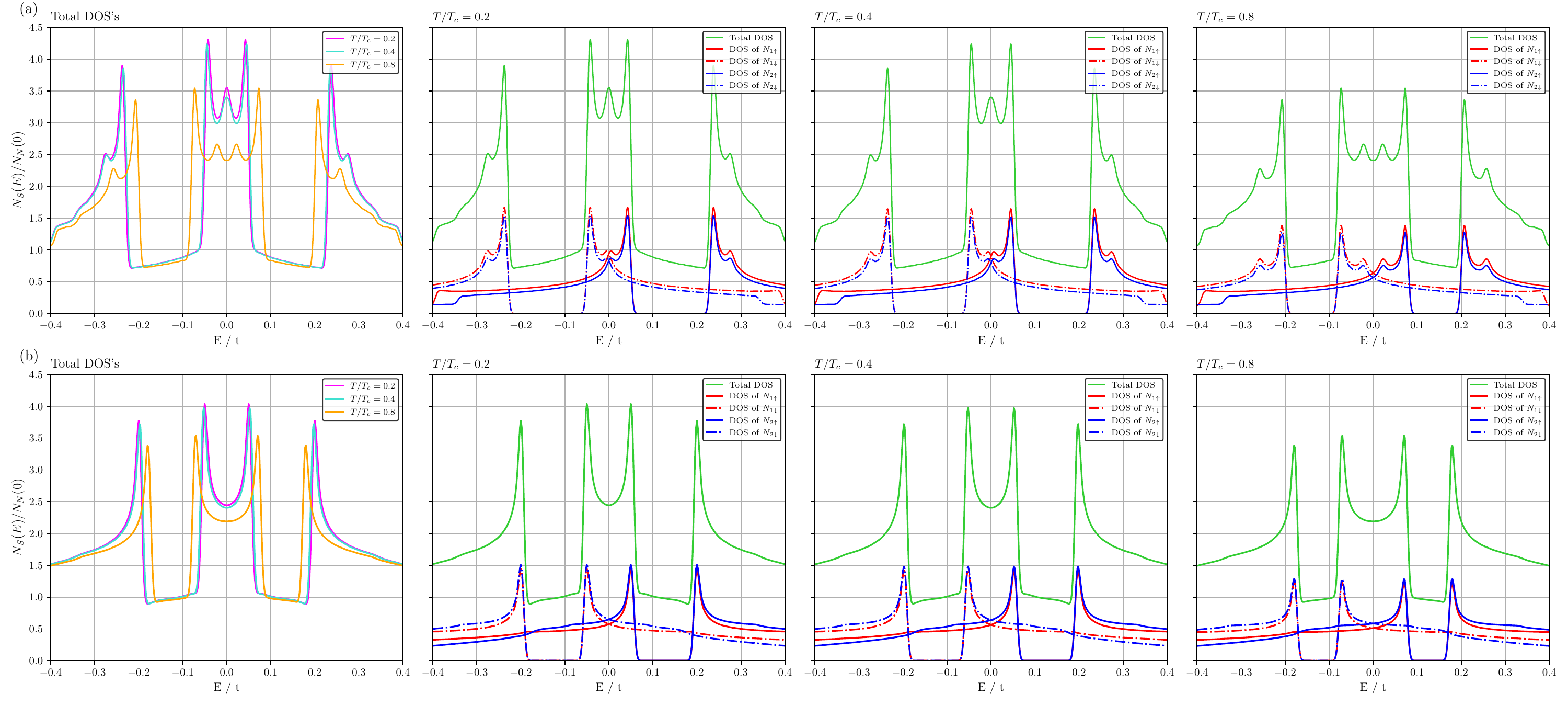}
 \caption{\label{fig:DOS2D}
Density of states $N_{S}(E;T,h)$ of the Mixing state 
for (a) 2D model with 2DFS2, $h/t=0.14$, and (b) 3D model with 3DFS2, $h/t=0.125$.}
\end{figure*}

We compute the DOS of the Bogoliubov quasiparticles as
\begin{align}
&N_{S}(E;T,h)=\sum_{m,\sigma} N_{m\sigma}(E;T,h) \\
&N_{m\sigma}(E;T,h)=\frac{1}{N^{d}}\sum_{\bm{k},s=\pm} 
\delta(E-(s E_{m\bm{k}}+(\hat{\sigma}_{z})_{\sigma \sigma}h)) \notag 
\end{align}
with an appropriate broadening introduced for numerical evaluation.
The results shown in Fig.~\ref{fig:DOS2D} 
clearly show that the Bogoliubov quasiparticle spectrum is gapless
in both the 2D and 3D cases. 
Remarkably, we find that the zero-energy DOS increases
as the temperature is lowered,
which is in sharp contrast to the behavior of conventional
fully gapped superconductors,
where the DOS is progressively suppressed at low temperatures.

Let us discuss the energy dependence of DOS in more detail. 
In the present parameter regime, the spin--orbit coupling and the inter-sublattice hybridization 
are sufficiently small that the quasiparticle spectrum is well approximated by 
$E_{m\bm{k}} \simeq \sqrt{\epsilon_{\bm{k}}^{2} + \Delta_{1}^{2}} .$
In this limit, the peak structure of the density of states is largely determined by $\epsilon_{\bm k}$. 
The condition $\epsilon_{\bm{k}}=0$, which defines the normal-state Fermi surface in the absence of SOC and hybridization, gives rise to the conventional BCS coherence peaks at $E=\pm \Delta_{1}+(\sigma_z)_{\sigma\sigma}h$.
In addition, the normal-state band possesses saddle points satisfying $\nabla_{\bm{k}}\epsilon_{\bm{k}}=0$, 
which produce van Hove singularities in two dimensions.
These saddle points are mapped to finite energies in the superconducting state 
%according to %$E_{vH} = \sqrt{\epsilon_{\mathrm{vH}}^{2} + \Delta_{1}^{2}}$, 
and therefore generate additional peak structures at 
$E = \pm\sqrt{\epsilon_{\mathrm{vH}}^{2} + \Delta_{1}^{2}}+(\sigma_z)_{\sigma\sigma}h$
in the Bogoliubov density of states.

Thus, the multiple peak structure observed in the 2D DOS can be understood as the combined effect of 
(i) the Fermi-surface contribution leading to the coherence peaks and 
(ii) the van Hove singularities of the underlying normal-state band.
In particular, the pronounced peak of the total DOS at $E=0$ in the 2D case at $T/T_c=0.2$ and $0.4$
(see Fig.~\ref{fig:DOS2D}(a))
originates from quasiparticle peaks associated with van Hove singularities in the underlying band structure.
In 3D, the additional momentum integration along $k_z$ significantly smears out 
the saddle-point singularities, and the corresponding peak structures 
are strongly reduced, consistent with our numerical results.

\begin{figure}
\includegraphics[width=\linewidth]{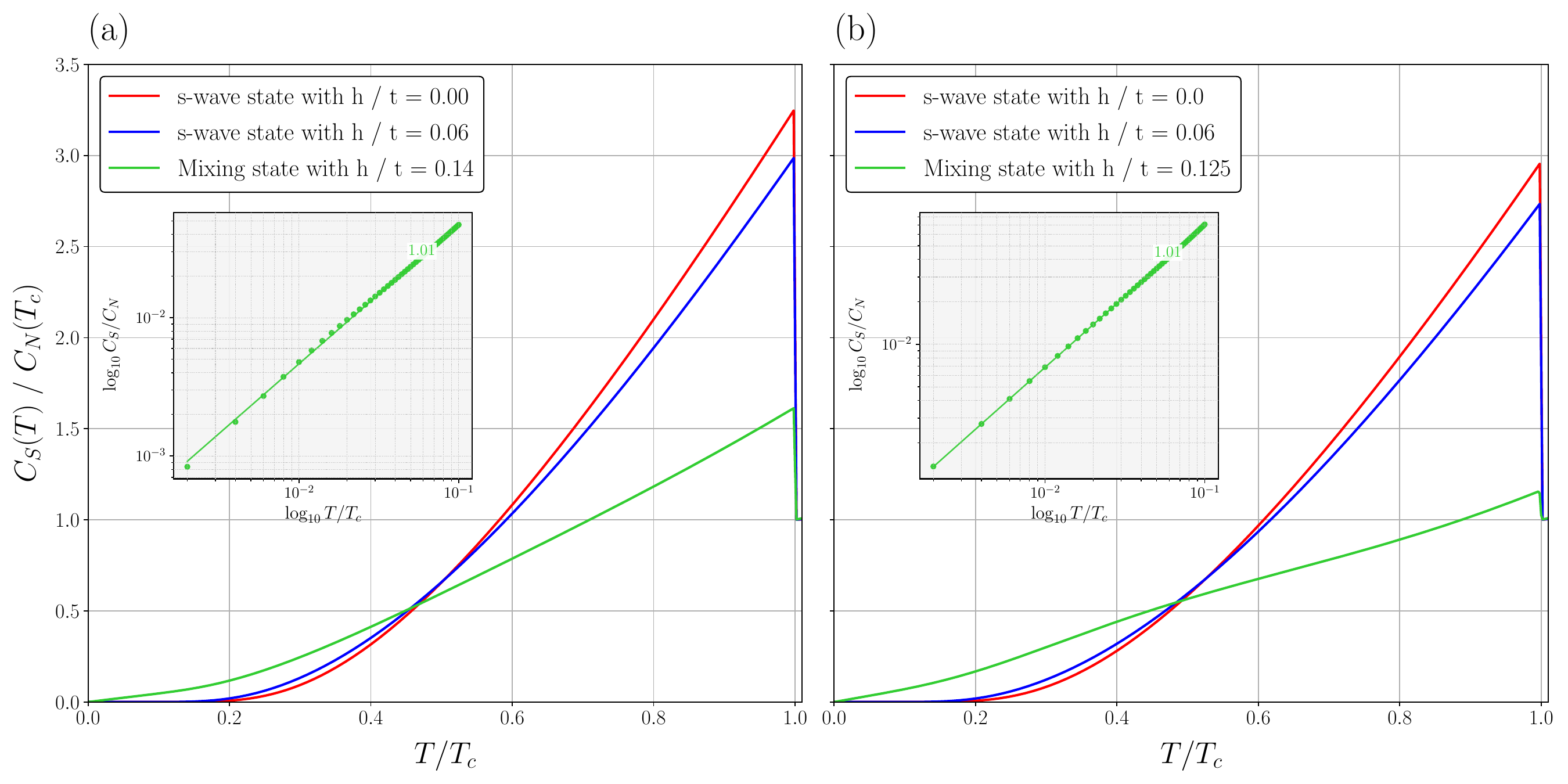}
\caption{\label{specific} 
Figure showing results of the specific heat for 
(a) 2D model with 2DFS2, and (b) 3D model with 3DFS2. 
Each inset in (a) and (b) shows the log-log plot 
for the Mixing state. }
\end{figure}

We numerically evaluate the specific heat $C_S(T)=-T \partial^2 F/ \partial T^2$ for both models, 
where $F=-k_B T \ln Z / N^d$ and $Z={\rm Tr}\left[ e^{-H/k_B T}\right]$.
The results in Fig.~\ref{specific} clearly demonstrate qualitative differences between 
the Mixing state and the conventional $s$-wave state. 
We can immediately see that $C_{S}(T)\propto T$ in Mixing state, reflecting 
the presence of finite zero-energy DOS (see Fig.~\ref{fig:DOS2D}, and insets in Fig.~\ref{specific}).

%%%%
%%%%

%%%%
%%%%

%========================= Conclusions part =========================
\section{Summary and Discussion}

In this work, we have studied interband pairing in a two-band superconductor
using minimal honeycomb tight-binding models with spin-orbit coupling and a Zeeman magnetic field.
We have shown that the superconducting state can be classified into two distinct types:
a conventional intraband $s$-wave pairing state and a Mixing state in which
intraband spin-triplet and interband spin-singlet pairing components coexist.
These two states are switched by an external magnetic field.

The Mixing state is stabilized at high magnetic fields,
where Zeeman splitting reduces the energy mismatch
between spin-split branches originating from different bands,
in agreement with previous studies \cite{highmag}.
Importantly, we show that a strong enhancement of the Mixing state
occurs when this Zeeman-induced near degeneracy takes place
at the Fermi level in the vicinity of a Van Hove singularity,
where the DOS is significantly amplified.
In contrast, the hybridization term $\epsilon_{c\bm{k}}$
and the spin-orbit coupling arising from local inversion-symmetry breaking
are essential for generating interband pairing components,
but do not enhance the Mixing state itself.

We have further demonstrated that the Mixing state remains stable
even in magnetic fields well above the Pauli-limiting field
and possesses a gapless Bogoliubov quasiparticle spectrum
in both 2D and 3D models.
As a consequence, the DOS is finite at zero energy
and, remarkably, increases as the temperature is lowered,
which is qualitatively opposite to the behavior of conventional
fully gapped superconductors.
This anomalous low-energy excitation structure gives rise to
distinct thermodynamic signatures,
most notably a $T$-linear dependence of the specific heat,
in sharp contrast to the activated behavior of a fully gapped
$s$-wave superconducting state.

We briefly comment on the relation to the Bogoliubov Fermi surface
proposed by Agterberg, Brydon, and Timm \cite{PhysRevLett.118.127001}.
In their work, a stable Bogoliubov Fermi surface arises
in multiband superconductors that spontaneously break
time-reversal symmetry in the pairing channel
and realize nonunitary pairing states,
for which the zero-energy manifold can be topologically protected.
In contrast, the Mixing state studied here
does not spontaneously break time-reversal symmetry.
Indeed, the gap matrix in Eq.~(\ref{mixingpair})
is invariant under the time-reversal transformation,
$\hat{\Delta}_{M\bm{k}}
\rightarrow
\hat{\sigma}_y \hat{\Delta}_{M-\bm{k}}^{*} \hat{\sigma}_y
=
\hat{\Delta}_{M\bm{k}}$.
Time-reversal symmetry is broken only explicitly
by the applied Zeeman field in our setup.
The gapless quasiparticle excitations originate from
Zeeman-induced band crossings and the branch structure
of the Bogoliubov spectrum.
While the resulting finite zero-energy DOS may resemble
that of a Bogoliubov Fermi surface,
it does not rely on the topological protection mechanism
identified in Ref.~\cite{PhysRevLett.118.127001},
but instead reflects the Zeeman-induced mechanism discussed above.

Because the quasiparticle excitation spectrum of the Mixing state
stabilized in a high magnetic field is intrinsically gapless,
a naive topological classification based on fully gapped bulk states
is not directly applicable.
Nevertheless, it may still be possible to discuss topological aspects
by introducing additional ingredients that remove the gapless nature
of the quasiparticle spectrum,
which we leave for future investigation.

The 2D honeycomb model considered here is directly relevant to
Zeeman-dominated monolayer superconductors,
such as graphene-based systems with enhanced spin-orbit coupling,
including graphene on WS$_2$ \cite{ws2} and stanene \cite{stanene},
where orbital pair-breaking effects can be strongly suppressed.

By contrast, the 3D non-symmorphic two-layer hexagonal model
should be interpreted as a charge-neutral fermionic superfluid
realized in an artificial lattice, such as an ultracold atomic system.
In such synthetic platforms, effective Zeeman fields can be implemented
without accompanying orbital effects,
making them a natural setting to explore the mechanism
for stabilizing interband pairing discussed in this work.
From this perspective, the 3D model serves
as a complementary platform to illustrate the generality of the mechanism,
rather than as a direct description of electronic superconductors.

We note that, in Zeeman-dominated superconductors under strong magnetic fields,
finite-center-of-mass-momentum pairing states such as the
Fulde--Ferrell--Larkin--Ovchinnikov (FFLO) state
may, in principle, compete with the spatially uniform superconducting phases
discussed in this work.
From a general viewpoint, such instabilities are naturally characterized
by the superfluid stiffness (phase rigidity),
which measures the stability of the uniform ($\bm{q}=0$) condensate
against phase twists or finite-momentum pairing.
Although the Mixing state hosts gapless quasiparticle excitations
that tend to suppress the superfluid stiffness,
the presence of a finite pairing amplitude $\Delta_1 \neq 0$
implies that the stiffness is expected to remain positive
in its stable regime,
indicating a genuine superconducting (superfluid) state.
A quantitative analysis of the superfluid stiffness and its possible sign change,
which would signal a transition to FFLO-type states,
requires a detailed treatment beyond the scope of the present paper
and is left for future investigation.
In this respect, the Mixing state shares certain similarities with the classical notion of gapless superconductivity discussed by Fulde and Maki~\cite{PhysRevLett.15.675}, where the quasiparticle gap vanishes while phase coherence remains intact.
However, unlike impurity-induced gapless states, the present mechanism originates from Zeeman-driven band crossings in a clean multiband system.

From a broader perspective, the Mixing state discussed here
provides a route to realizing unconventional, gapless superfluid behavior
without invoking exotic pairing mechanisms.
Our findings highlight the potential of magnetic-field control,
combined with multiband structure and enhanced DOS,
as a powerful tool to engineer interband pairing
and novel thermodynamic responses
in both solid-state and synthetic quantum systems.

\begin{acknowledgments}
 
 The authors thank H. Ueki for fruitful discussions. 
 
\end{acknowledgments}

\bibliographystyle{apsrev4-2}
\bibliography{draft}

\vspace{1cm}

%\newpage

\end{document}